\documentclass[final,1p,times]{elsarticle} 
\usepackage{graphicx} 
\usepackage{amssymb} 
\usepackage{amsthm} 
\usepackage{lineno} 
\usepackage{epstopdf}
\journal{Nuclear Physics A} 

\newcommand{\sNN}[1]{$\sqrt{s_{NN}} = #1$ GeV}

\begin{document} 

\begin{frontmatter} 


\title{$D^0$ measurements in Au+Au Collisions at \sNN{200}  
using the  STAR Silicon Inner Tracker}

\author{Sarah LaPointe for the STAR Collaboration}

\address{Department of Physics and Astronomy, Wayne State University, 666 W. Hancock, Detroit, MI 48201, USA}

\begin{abstract} 
We present preliminary results from $D^{0}$ meson measurements through the hadronic decay channel in minimum bias Au+Au collisions at \sNN{200} at STAR. The measurements are performed 
using a secondary vertexing technique that exploits the resolution given by the Silicon 
detectors available in STAR. 
\end{abstract} 

\end{frontmatter} 



\section{Motivation}

Charm quarks, because of their large masses, are expected to be produced predominately from the initial gluon fusion in parton-parton hard scatterings\cite{fusion}. This indicates that the production of the charm occurs at the early stages of the collision. At this time the system is thought to be partonic, therefore the charm can be used as a powerful probe of the initial conditions. A direct reconstruction of the $D^{0}$ through its 
hadronic decay channel will yield a less ambiguous measurement of $R_{AA}$, the nuclear 
modification factor, and $\nu_2$, elliptical flow than the recently published semi-leptonic decay measurements from the RHIC experiments\cite{star_e, phenix_e}. Such measurements gives insight as to how charm quarks interact with the medium. 

\section{Procedure and Results}

The $D^{0}$ analysis is based on data taken with the STAR Time Projection Chamber (TPC), 
Silicon Vertex Tracker (SVT), and Silicon Strip Detector (SSD). Tracking is performed 
using all three of the detectors. Fig. 1 shows that at $p_T$ = 1 GeV/c the inclusion of 2 or more hits from the SVT and SSD improves the impact parameter resolution of the global tracks to the primary vertex by 
more than a factor of 10 compared to TPC alone. In addition to tracking, the TPC also provides
information on the energy loss (dE/dx) of the charged tracks, which is used as a tool for 
particle identification.

 \begin{figure}[ht]
\centering
\includegraphics[scale=0.4]{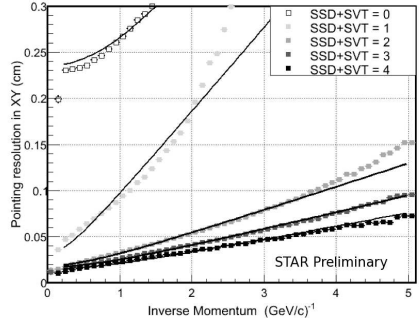}
\caption{The global track impact parameter (pointing) resolution in the transverse plane as a function of 1/$p_{T}$ with various numbers of silicon hits\cite{nim}.}
\label{Fig1} 
\end{figure}

The displaced decay vertex of the $D^{0}$ (c$\tau$=123 $\mu$m) occurs before the tracking detectors.  
Thus, all oppositely charged identified tracks are paired and projected toward the primary 
collision vertex. If the two trajectories cross at some point before the primary vertex, they 
are considered candidates for the $D^{0}$ decay. The $D^{0}$ is identified through an invariant 
mass analysis of its decay to $K^{-}$$\pi^{+}$ after the application of a set of geometric cuts which 
discriminate between the signal and the background.
%

%

The analysis was performed using 17 million minimum bias Au+Au collisions at $\sqrt{s_{NN}}$ = 
200 GeV. Fig. 2 (left) shows the invariant mass obtained after identifying kaons and 
pions using dE/dx, requiring both daughter tracks to have 3 SVT hits, and applying 
geometrical variable cuts. At this stage the $D^{0}$ peak is not easily visible, so the 
background is subtracted by fitting a polynomial to the Ôside bandsÕ of the 
distribution (see Fig. 2(left)) and subtracting the background under the peak 
based on the $\chi^{2}$ minimized polynomial fit. The preliminary result is Figure 2 (right). The 
signal yield is $\simeq$ 3000 with a significance S/$\sqrt{S+B}$ $\simeq$ 4.5, and a S/B of $\sim$ 0.007. 
\begin{figure}[h]
\begin{tabular}{c c}
\includegraphics[width = 0.45\textwidth]{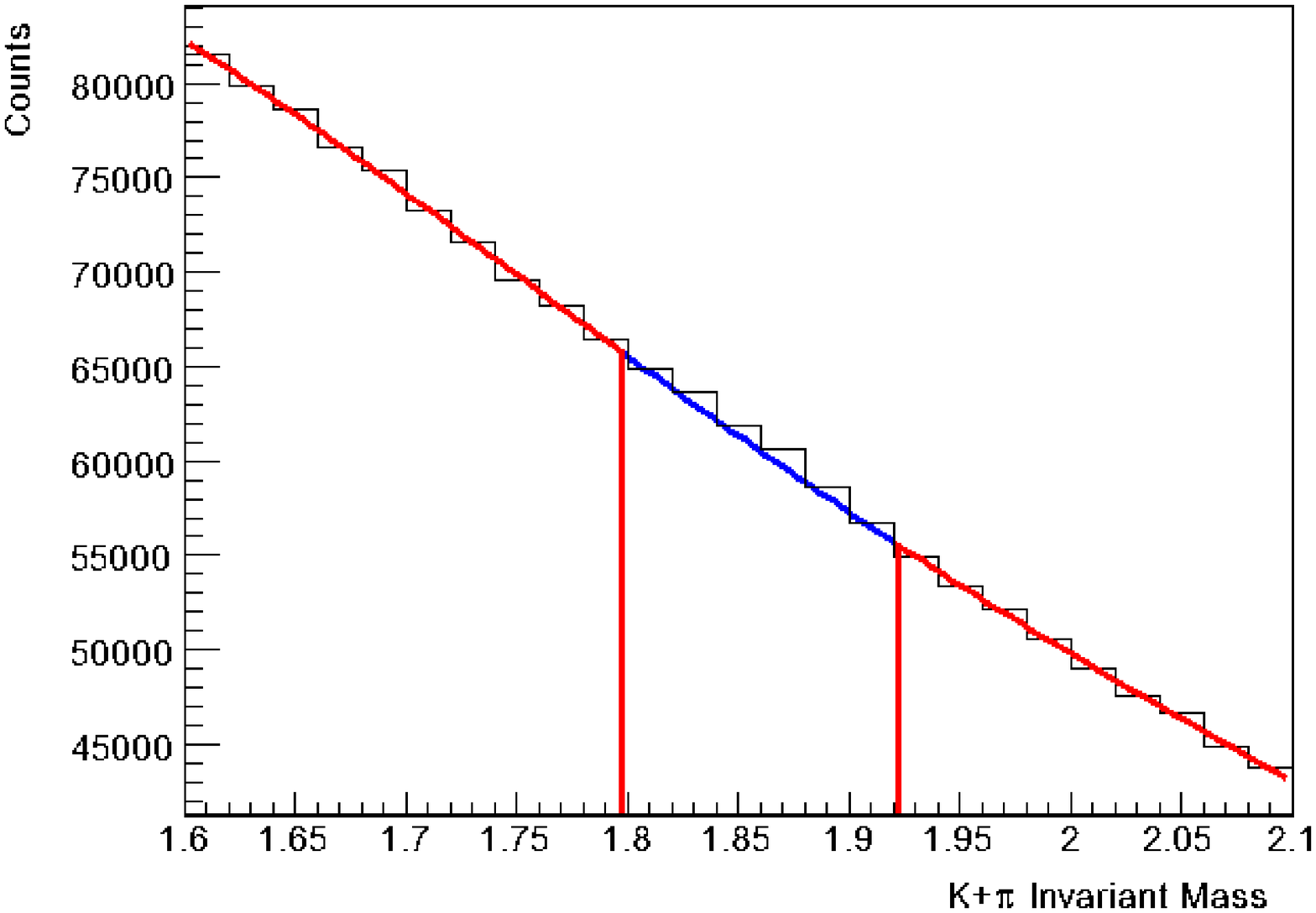}
&
\includegraphics[width = 0.45\textwidth]{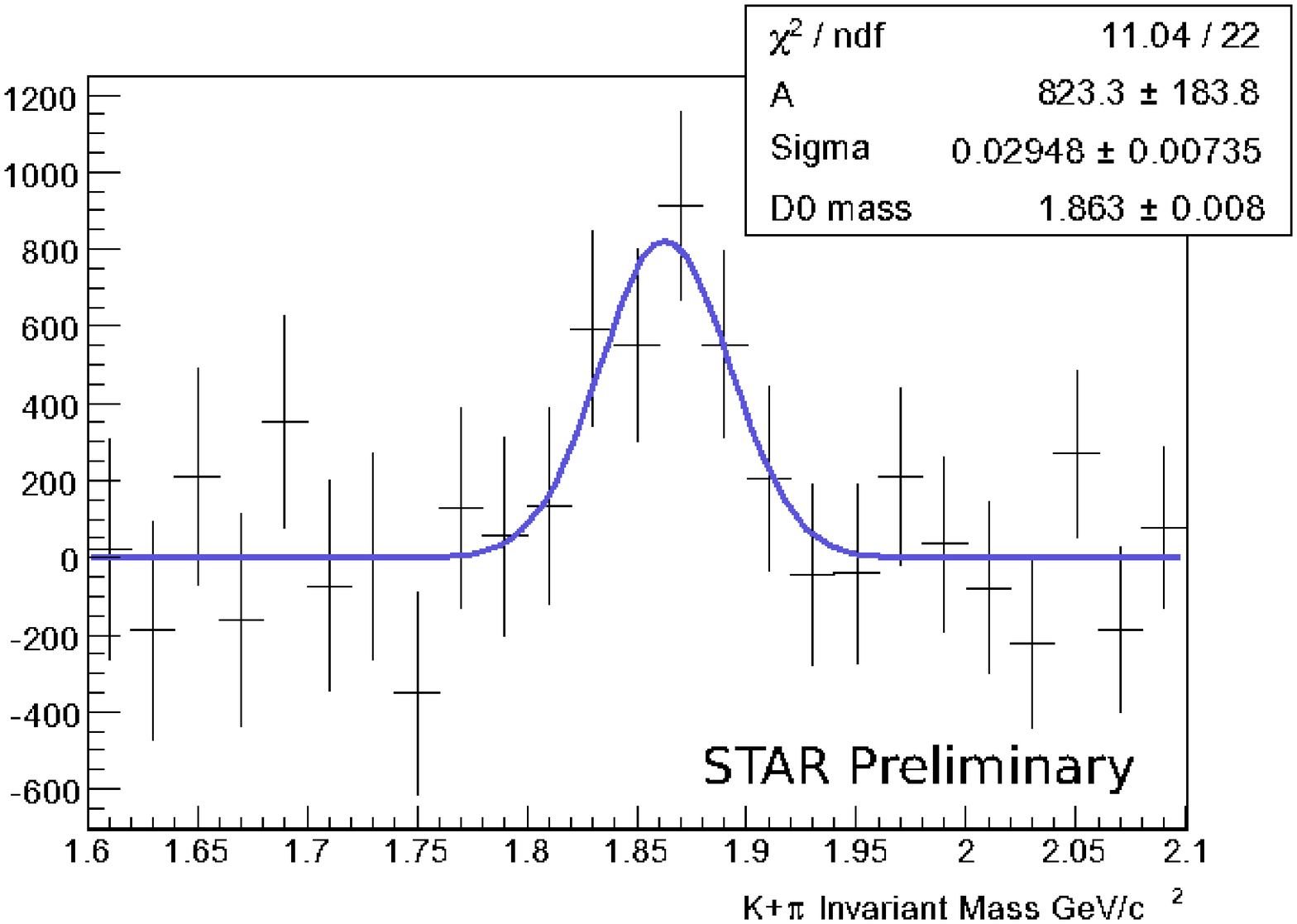}
\end{tabular}
\caption{Left: K$\pi$ invariant mass after optimized cuts Right: K$\pi$ invariant mass after 
background subtraction}
\label{Fig2}
\end{figure} 
\section{Outlook}
With an additional 30 million events that still need to be analyzed, and further optimization of the geometrical cuts, we estimate a direct measurement of D mesons out to a $p_{T}$ of 4GeV/c, from which we expect to measure a total charm cross section in NN collisions, and to attempt a measurement of the D meson $R_{CP}$. 


\end{document}